\documentclass[letterpaper]{article} 
\usepackage{aaai25}  
\usepackage{times}  
\usepackage{helvet}  
\usepackage{courier}  
\usepackage[hyphens]{url}  
\usepackage{graphicx} 
\usepackage{natbib}  
\usepackage{caption} 
\usepackage{algorithm}
\usepackage{algorithmic}
\usepackage{amsmath}
\usepackage{amssymb}
\usepackage[capitalize]{cleveref}
\usepackage{color}
\usepackage{newfloat}
\usepackage{listings}
\usepackage{multirow}
\usepackage{marvosym}
\DeclareCaptionStyle{ruled}{labelfont=normalfont,labelsep=colon,strut=off} 
\floatstyle{ruled}
\newfloat{listing}{tb}{lst}{}
\floatname{listing}{Listing}
\newcommand{\eg}{\emph{e.g.}}

\definecolor{gblack}{RGB}{80,80,80}

\title{ProGDF: Progressive Gaussian Differential Field for \\ Controllable and Flexible 3D Editing}

\author{
    Yian Zhao\textsuperscript{1,2} \quad Wanshi Xu\textsuperscript{1,2} \quad Yang Wu\textsuperscript{2\ \Letter} \quad Weiheng Huang\textsuperscript{2} \quad Zhongqian Sun\textsuperscript{2} \quad Wei Yang\textsuperscript{2}
}
\affiliations{
    \textsuperscript{\rm 1} Peking University \quad \textsuperscript{\rm 2} Tencent AI Lab \\
}

\usepackage{bibentry}

\begin{document}

\maketitle

\begin{abstract}
3D editing plays a crucial role in editing and reusing existing 3D assets, thereby enhancing productivity.
Recently, 3DGS-based methods have gained increasing attention due to their efficient rendering and flexibility.
However, achieving desired 3D editing results often requires multiple adjustments in an iterative loop, resulting in tens of minutes of training time cost for each attempt and a cumbersome trial-and-error cycle for users.
This in-the-loop training paradigm results in a poor user experience.
To address this issue, we introduce the concept of process-oriented modelling for 3D editing and propose the Progressive Gaussian Differential Field~(\textbf{ProGDF}), an out-of-loop training approach that requires only a single training session to provide users with controllable editing capability and variable editing results through a user-friendly interface in real-time.
ProGDF consists of two key components: Progressive Gaussian Splatting~(\textbf{PGS}) and Gaussian Differential Field~(\textbf{GDF}).
PGS introduces the progressive constraint to extract the diverse intermediate results of the editing process and employs rendering quality regularization to improve the quality of these results.
Based on these intermediate results, GDF leverages a lightweight neural network to model the editing process.
Extensive results on two novel applications, namely controllable 3D editing and flexible fine-grained 3D manipulation, demonstrate the effectiveness, practicality and flexibility of the proposed ProGDF.
\end{abstract}
\footnote{\Letter\ Corresponding author.}
\section{Introduction}
3D digital assets play a crucial role in various high-level applications, \eg, virtual reality, gaming, and the movie industry. However, the traditional pipeline for creating 3D assets is both expensive and time-consuming, requiring significant labor investment at each step. Recently, NeRF~\cite{mildenhall2021nerf} and 3DGS~\cite{kerbl20233d} have gained attention for their superior modeling quality and differentiable rendering pipeline. Building upon these advancements, several works~\cite{poole2022dreamfusion,wang2023score} utilize pre-trained diffusion model~\cite{rombach2022high} to achieve efficient 3D generation. However, the generated 3D assets may not fully meet the user's requirements in terms of geometry and appearance. Therefore, the development of user-friendly 3D editing methods becomes indispensable.

\begin{figure}[t]
\hsize=\linewidth
\centering
\includegraphics[width=\linewidth]{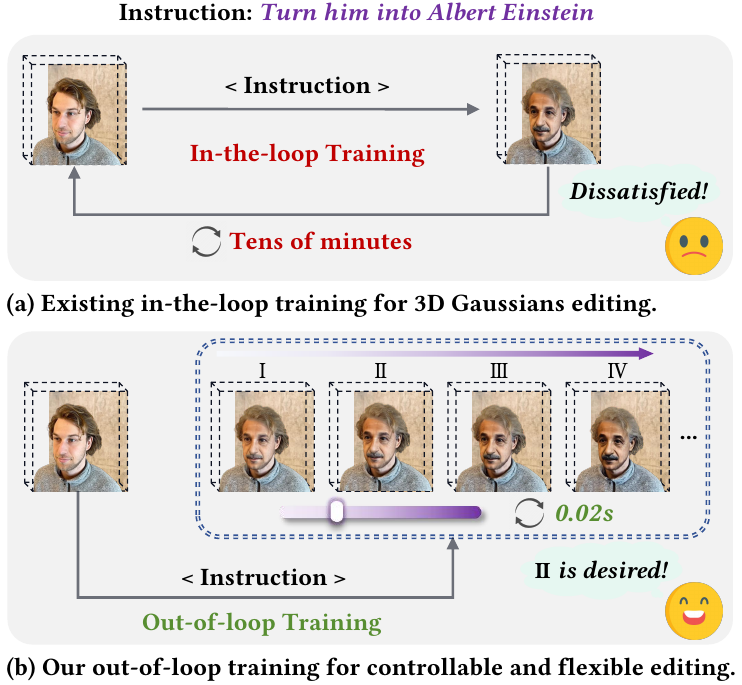}
\caption{
\textbf{(a)}: Existing in-the-loop training for 3D Gaussians editing, takes tens of minutes for each loop, resulting in a cumbersome trial-and-error cycle for users.
\textbf{(b)}: Our out-of-loop training approach, requiring only a single training session to provide controllable editing capability and variable editing results through a user-friendly interface, with each adjustment taking only $0.02$ seconds.
}
\label{fig:motivation}
\end{figure}
\begin{figure*}[ht]
\hsize=\linewidth
\centering
\includegraphics[width=\linewidth]{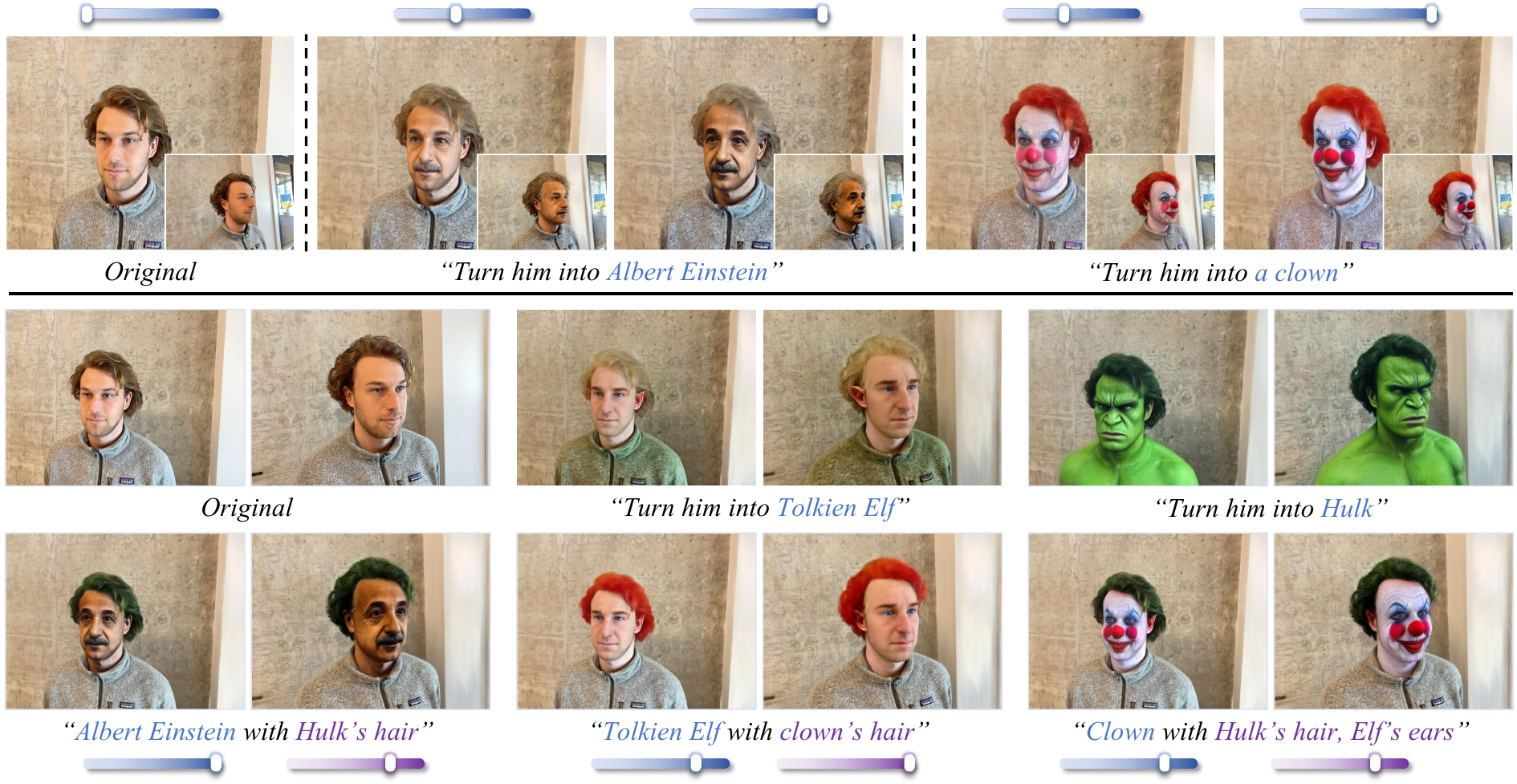}
\caption{
{Results of ProGDF.} Our ProGDF not only achieves controllable 3D editing with a user-friendly interface to generate diverse results at varying levels of editing~(see first row), but also enables flexible fine-grained 3D manipulation to create new results from existing edited 3D scenes~(see second row). 
Note that the instructions marked in the second row are not real inputs, but are used to indicate which editing results they are combined from.
}
\label{fig:front_result}
\end{figure*}

Traditional 3D editing methods typically rely on meshes and point clouds as manipulation objects. However, the inherent complexity of these representations poses challenges in developing user-friendly interfaces and rendering complex 3D scenes. With the emergence of implicit neural radiance field~(NeRF), several NeRF-based 3D editing frameworks~\cite{liu2021editing,haque2023instruct} have been proposed. Nevertheless, the applicability of NeRF is limited by its reliance on the Multi-Layer Perception (MLP) network for scene data encoding. This restricts direct modification of specific parts and complicates the task, including the training and rendering processes, hindering practical applications.

Recently, the efficiency and explicit nature of 3DGS have made it easier to manipulate specific regions, demonstrating a distinct advantage in 3D editing. Existing methods~\cite{chen2024gaussianeditor,wang2024gaussianeditor} achieve impressive results by optimizing specific 3D Gaussians using image editing models~\cite{brooks2023instructpix2pix}. 
However, achieving desired 3D editing results often requires multiple adjustments in an iterative loop. Existing methods typically involve training numerous 3D Gaussians for each editing session, resulting in tens of minutes of training time cost for each attempt and a cumbersome trial-and-error cycle of adjusting hyper-parameters.
This in-the-loop training paradigm results in a poor user experience, \textit{cf.}~\cref{fig:motivation}(a).

To address this issue, we introduce the concept of process-oriented modelling for 3D editing and propose the Progressive Gaussian Differential Field~(\textbf{ProGDF}), an out-of-loop training approach that requires only a single training session to provide users with controllable editing capability and variable editing results through a user-friendly interface in real-time, eliminating the need for complex hyper-parameters adjustments and retraining in a new session, \textit{cf.}~\cref{fig:motivation}(b). 
Compared to previous methods, our approach presents the following advantages: (\romannumeral1). \textit{Real-time feedback}, with each adjustment taking just $0.02$ seconds, (\romannumeral2). \textit{Controllable 3D editing}, providing users with controllable editing capability and variable editing results through a interactive interface, and (\romannumeral3). \textit{Flexible fine-grained 3D manipulation}, supporting independent control of the editing of 3D Gaussians at any position.
Some representative results are presented in \cref{fig:front_result}.

The proposed ProGDF consists of two key components: Progressive Gaussian Splatting~(\textbf{PGS}) and Gaussian Differential Field~(\textbf{GDF}). PGS introduces progressive constraint and rendering quality regularization into the original 3DGS. The former extracts diverse intermediate results by tracking the optimization trajectory of 3D Gaussians from the original scene to the target scene, while the latter improves the quality of these intermediate results. 
Based on these results, GDF, consisting of a lightweight MLP, is designed to model the 3D editing process as an estimation of the difference between the original and edited 3D Gaussians, and provides an interactive interface to control the editing results, offering the user a variety of choices, thus increasing the editing success rate and improving the user experience.

\begin{figure*}[ht]
\hsize=\linewidth
\centering
\includegraphics[width=0.98\linewidth]{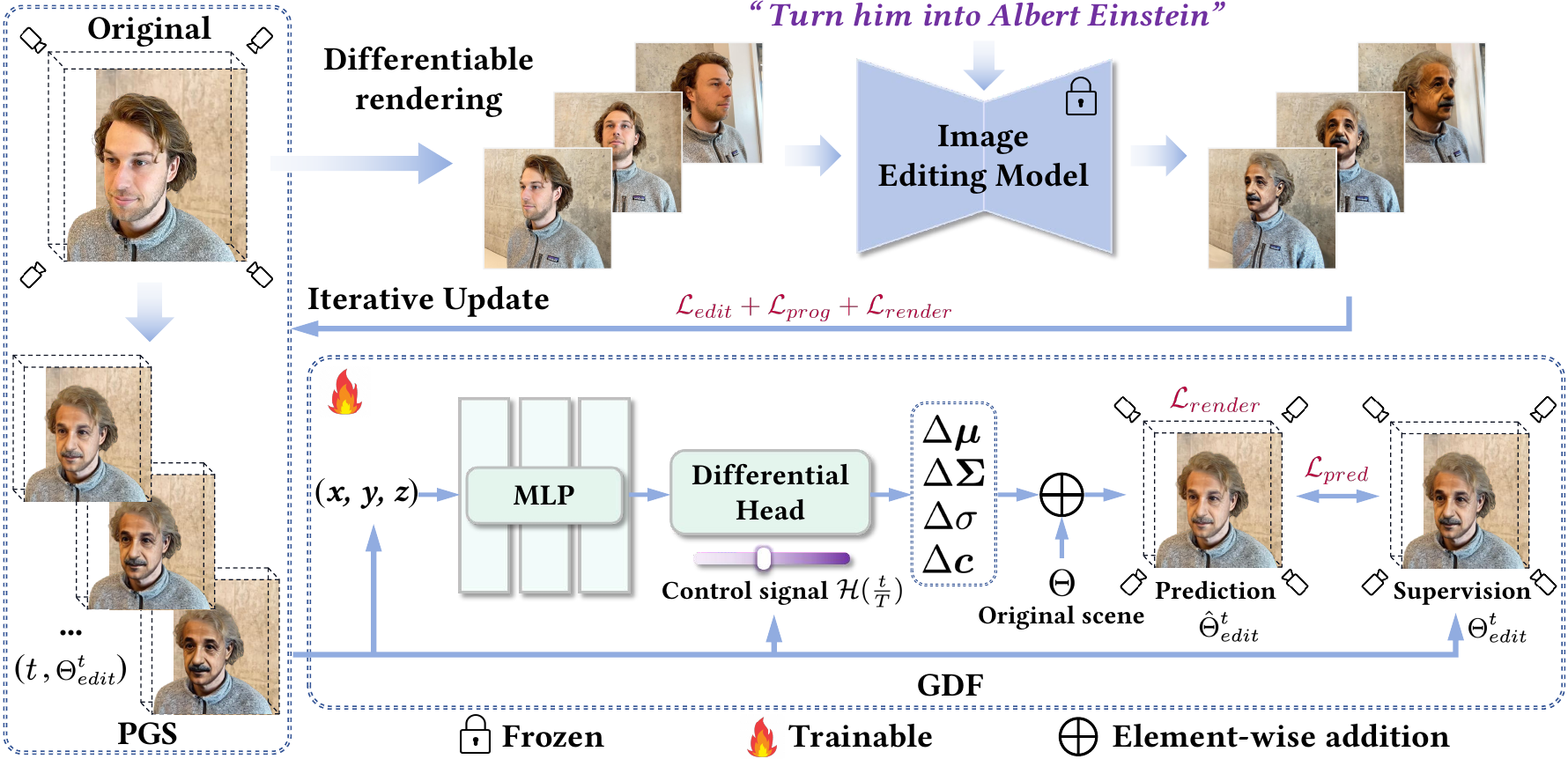}
\caption{
{Overview of ProGDF.} ProGDF contains two key components: Progressive Gaussian Splatting~(PGS) and Gaussian Differential Field~(GDF).
PGS is utilized to extract diverse high-quality intermediate results by tracking the optimization trajectory of 3D Gaussians from the original scene to the target scene.
GDF leverages a lightweight MLP to model the 3D editing process as an estimation of the difference between the original and the edited 3D Gaussians, providing controllable editing capability and variable editing results through a user-friendly interface in real-time.
}
\label{fig:overview}
\end{figure*}

We perform experiments on multiple 3D scenes from multiple datasets.
Extensive results on two novel applications, namely controllable 3D editing and flexible fine-grained 3D manipulation, fully demonstrate the effectiveness, practicality, and flexibility of the proposed ProGDF.
The main contributions of this work can be summarized as follows: (\romannumeral1). opening up an out-of-loop training approach to replace the existing in-the-loop training paradigm for 3D editing, eliminating the need for adjusting complex hyper-parameters and retraining within the editing loop, (\romannumeral2). introducing the Progressive Gaussian Differential Field to achieve the process-oriented modeling for 3D editing, providing users with controllable editing capability and variable editing results through a user-friendly interface in real-time, and (\romannumeral3). proposing two novel applications, namely controllable 3D editing and flexible fine-grained 3D manipulation, to extend the controllability and flexibility of the 3D editing task.

\section{Methodology}

\subsection{Overall Pipeline}
In this section, we detail the proposed Progressive Gaussian Differential Field~(ProGDF), \textit{cf.}~\cref{fig:overview}, which includes two key components: Progressive Gaussian Splatting~(PGS) and Gaussian Differential Field~(GDF).
Formally, Gaussian Splatting~\cite{kerbl20233d} explicitly represents a 3D scene as a set of 3D Gaussian primitives $ \Theta = \{ (\boldsymbol{\mu}_i, \boldsymbol{\Sigma}_i, \sigma_i, \boldsymbol{c}_i) \}_{i=1}^N $.
Given an editing instruction $e$ and a 3D scene represented by 3D Gaussians $ \Theta$, we first employ the image editing model to process multi-view images according to $e$, and then optimize the 3D Gaussians to obtain the edited scene $\Theta_{edit}$.
Meanwhile, the GDF takes a position in space as input and estimates the offset of the properties of 3D Gaussians at that position due to editing.
The PGS extracts the high-quality intermediate results on the optimization trajectory of 3D Gaussians and provides diverse supervision for the GDF, and the control signals computed from the time steps of the trajectory afford editing controllability.
For training, the GDF is lightweight and can be optimized in parallel with 3D Gaussians, resulting in marginal time and memory consumption, and importantly we only need to train it once.
After a single out-of-loop training session, GDF is able to deliver controllable editing capability and variable editing results in $0.02$ seconds via a user-friendly interface.

\subsection{Progressive Gaussian Splatting}
\noindent \textbf{3D Editing From 2D Priors.}
Given the original scene $\Theta$, we first obtain the multi-view image set $\boldsymbol{I}_r$ using the differentiable renderer proposed by~\cite{kerbl20233d}~(see the Appendix for more details), and then employ the image editing model $\mathcal{E}$ to edit the images in $\boldsymbol{I}_r$ according to the instruction $e$ to obtain the edited image set $\boldsymbol{I}_r^e$, \textit{cf.}~\cref{eq:edit}.
\begin{equation}
\label{eq:edit}
    \boldsymbol{I}_r^e = \{ \mathcal{I}_r^e \ | \ \mathcal{I}_r^e = \mathcal{E}(\mathcal{I}_r, e), \ \mathcal{I}_r \in \boldsymbol{I}_r \}, 
\end{equation}
To edit the 3D Gaussians, we calculate the editing loss between images within $\boldsymbol{I}_r$ and $\boldsymbol{I}_r^e$ to optimize the parameters of 3D Gaussians.
Since the input images of the 2D editing model $\mathcal{E}$ are rendered from the original 3D scene, the viewpoint-independent constraint implicitly encompasses the intrinsic 3D consistency.
To maintain the background consistency, we utilize SAM~\cite{kirillov2023segment} to segment multi-view images and unproject the 2D masks into 3D scene follow~\cite{chen2024gaussianeditor}, editing only the 3D Gaussians contained in the selected region~(see the Appendix for calculation details).
The editing loss function as in~\cref{eq:edit_loss}.
\begin{equation}
\label{eq:edit_loss}
\begin{aligned}
    \mathcal{I}_r^t = \mathcal{R}(\Theta^t, v), \ &1 \leq t \leq T, \\
    \mathcal{L}_{edit} = \mathcal{L}_1(\mathcal{I}_r^t,\mathcal{I}_r^e) +& \mathcal{L}_{lpips}(\mathcal{I}_r^t, \mathcal{I}_r^e), 
\end{aligned}
\end{equation}
where $\mathcal{R}$ represents the differentiable renderer, $\Theta^t$ represents the 3D Gaussians at time step $t$, $v$ is the viewpoint of $\mathcal{I}_r^e$, $T$ is the total steps, and $\mathcal{L}_1$ and $\mathcal{L}_{lpips}$ are utilized to calculate the loss between images.

\noindent \textbf{Progressive Constraint.}
To enable process-oriented modelling for 3D editing, only the editing loss $\mathcal{L}_{edit}$ is not enough.
We observe that in the early steps of editing, a large number of 3D Gaussians are updated in an imprecise direction, resulting in a tendency to blur textures and distort details in the scene.
As training progresses, the 3D scene converges from chaos to the target scene, which restricts the ability to extract diverse intermediate results along the optimization trajectory.
To solve this issue, we propose the progressive constraint to smooth the optimization trajectory by adaptively weighting the update magnitude of 3D Gaussians in each iteration.
Specifically, we regard 3D editing as a progressive approach from the original scene $\Theta$ to the target scene $\Theta_{edit}$, in which the update magnitude of 3D Gaussians decays progressively and finally converges to $\Theta_{edit}$.
We denote the initial weight as $\alpha$, which decays step-wise along the optimization trajectory of 3D Gaussians.
The progressive constraint is shown in~\cref{eq:prog}.
\begin{equation}
\label{eq:prog}
    \mathcal{L}_{prog} = \alpha\cdot\beta^{\frac{t}{s}}\cdot\sum_{i=1}^{N}\Delta\Theta_i^t, \ 1 \leq t \leq T,
\end{equation}
where $\beta$ is the decay coefficient, $\Delta \Theta_i^t$ represents the variation of $i$-th 3D Gaussian at time step $t$, and $s$ serves to mitigate the exponential explosion.

\noindent \textbf{Rendering Quality Regularization.}
In addition, the progressive constraint does not explicitly constrain the texture blurring and detail distortion for intermediate results, resulting in poor quality supervision for GDF.
To address this issue, we propose the rendering quality regularization.
We employ the Laplacian gradient~\cite{wang2007laplacian} of the rendered image to evaluate its quality.
In general, the greater the sharpness of an image, the larger its gradient, and vice versa.
Experiments demonstrate that applying the simple regularization yields impressive results. The rendering quality regularization is shown in~\cref{eq:quality}.
\begin{equation}
\label{eq:quality}
    \mathcal{L}_{render} = -\nabla_{Laplacian}^2 \mathcal{R}(\Theta^t, v), \ 1 \leq t \leq T,
\end{equation}
The loss function for PGS is a weighted sum of $\mathcal{L}_{edit}$, $\mathcal{L}_{prog}$ and $\mathcal{L}_{render}$ as in~\cref{eq:loss_pgs}. 
\begin{equation}
\label{eq:loss_pgs}
    \mathcal{L}_{PGS} = \lambda_1\mathcal{L}_{edit} + \lambda_2\mathcal{L}_{prog} + \lambda_3\mathcal{L}_{render},
\end{equation}
where $\lambda_1$, $\lambda_2$, $\lambda_3$ are taken as $1$, $5$, $1$, respectively.

\begin{figure*}[ht]
\hsize=\linewidth
\centering
\includegraphics[width=\linewidth]{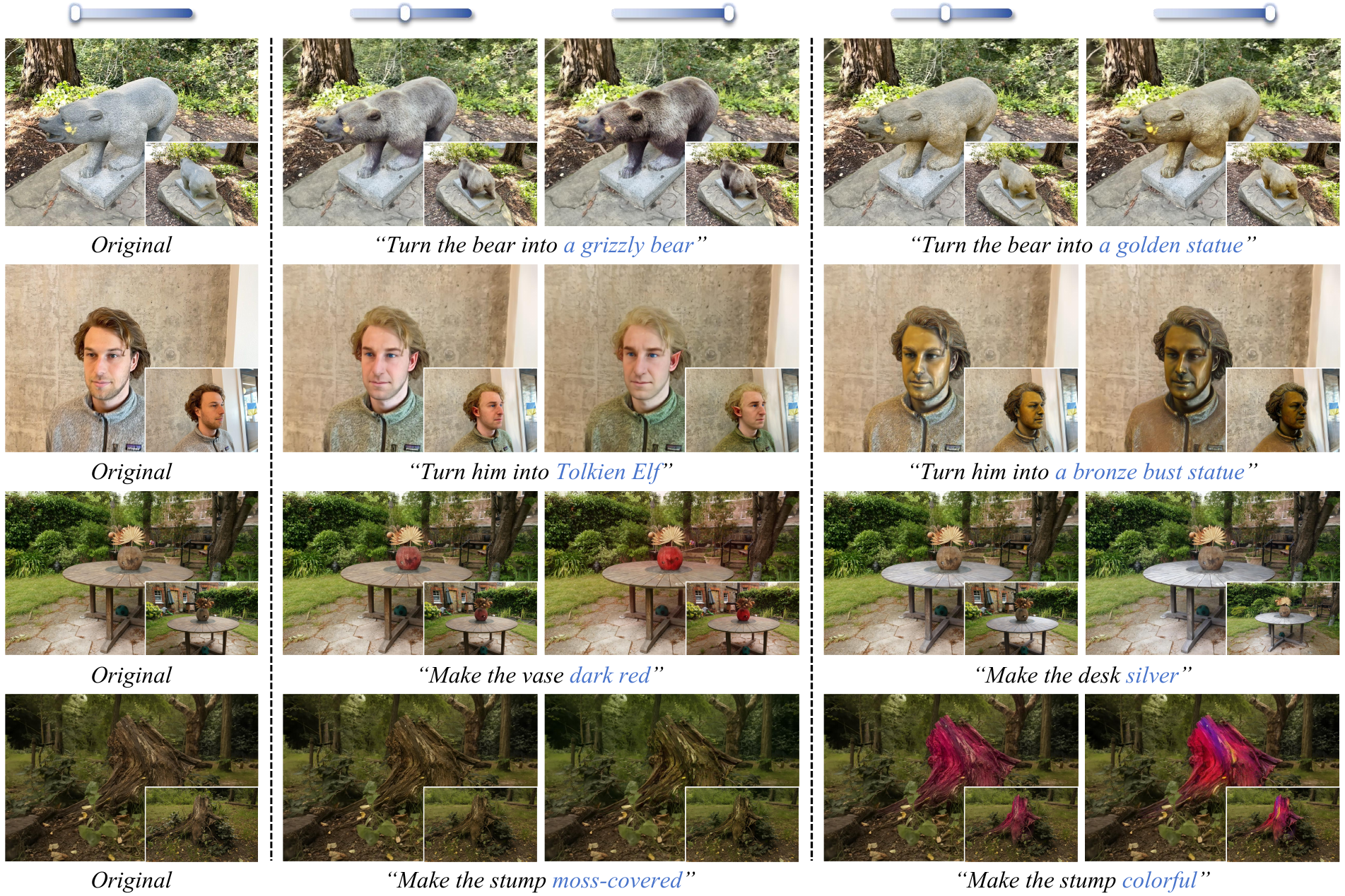}
\caption{
{Results of controllable 3D editing.}
Our method is capable of editing a variety of scenes. Only the 3D Gaussians within the region to be edited are inputted into the GDF, leaving the other region unaffected. We present two editing instructions for each scene, with two different results for each instruction, to demonstrate the controllability of the editing.
}
\label{fig:controllable}
\end{figure*}
\begin{figure*}[!ht]
\hsize=\linewidth
\centering
\includegraphics[width=\linewidth]{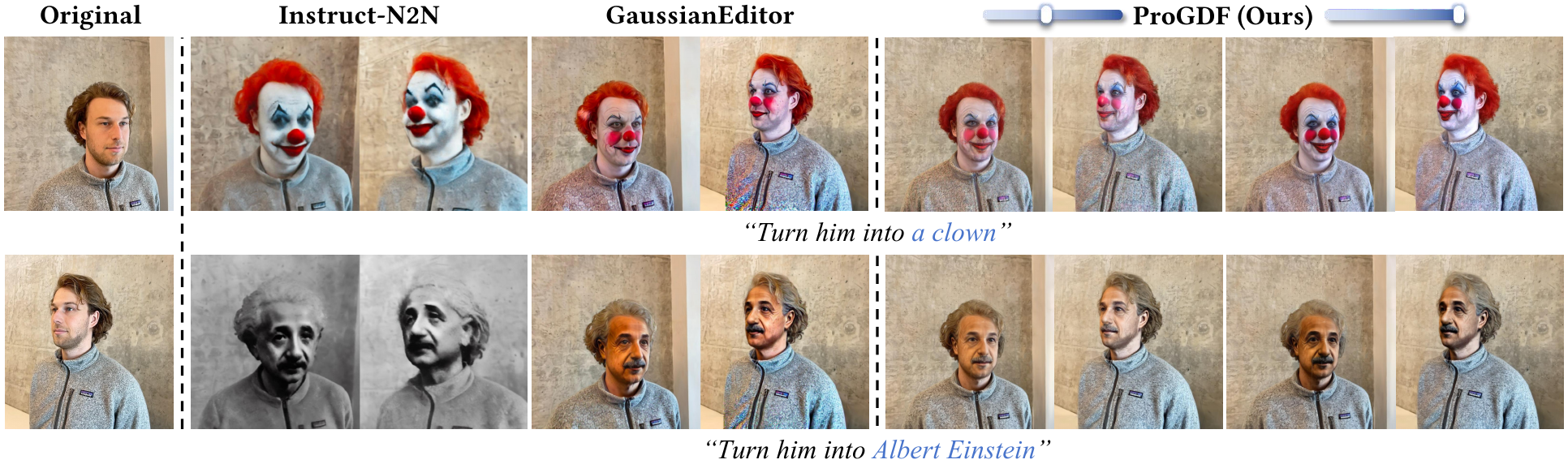}
\caption{
{Comparison with previous methods.} We compare our method with Instruct-N2N~\cite{haque2023instruct}~(NeRF-based) and GaussianEditor~\cite{chen2024gaussianeditor}~(3DGS-based), respectively. 
Neither allows for controllable editing, whereas our method is capable of controlling the editing result through a user-friendly interface.
}
\label{fig:comparison}
\end{figure*}

\subsection{Gaussian Differential Field}

\noindent \textbf{Framework Design.}
The proposed PGS opens up the possibility of process-oriented modelling by tracking the optimization trajectory of 3D Gaussians and providing diverse high-quality intermediate results of the editing process.
However, the number of parameters for these intermediate results is large and not visible to users.
To present these results in a user-friendly manner, we parameterize the 3D editing process from the original 3D scene $\Theta$ to the target 3D scene $\Theta_{edit}$ with a lightweight neural network called Gaussian Differential Field~(GDF), denoted as $\mathcal{G}$.
In detail, the input to the GDF is any point $(x, y, z)$ in space and the output is the offset of the 3D Gaussian $(\Delta \boldsymbol{\mu}_i, \Delta \boldsymbol{\Sigma}_i, \Delta \sigma_i, \Delta \boldsymbol{c}_i)$ at that position due to editing, which is added to the original 3D Gaussian $\Theta_i$ to obtain the edited results.
Note that we ignore points in space where no 3D Gaussian exists and utilize $(\Delta \boldsymbol{S}, \Delta \boldsymbol{R})$ instead of $\Delta \boldsymbol{\Sigma}$ in the implementation. 
We utilize $\boldsymbol{\mathcal{M}}$ to represent the 3D mask, where the 3D Gaussians to be edited are marked with $1$ and the others with $0$. 
The predicted edited 3D Gaussians $\hat{\Theta}_{edit}$ are calculated according to~\cref{eq:gdf1} and \cref{eq:gdf2}.
\begin{small}
\begin{equation}
\label{eq:gdf1}
(\Delta \boldsymbol{\mu}_i, \Delta \boldsymbol{\Sigma}_i, \Delta \sigma_i, \Delta \boldsymbol{c}_i) =
\begin{cases}
    {\mathcal{G}}(\boldsymbol{\mu}_i), & \boldsymbol{\mathcal{M}}_i = 1, \\
    \overrightarrow{\mathbf{0}}, & \boldsymbol{\mathcal{M}}_i = 0,
\end{cases}
\end{equation}
\end{small}
\begin{small}
\begin{equation}
\label{eq:gdf2}
    \hat{\Theta}_{edit} = \{ \boldsymbol{\mu}_i + \Delta \boldsymbol{\mu}_i, \boldsymbol{\Sigma}_i + \Delta \boldsymbol{\Sigma}_i, \sigma_i + \Delta \sigma_i, \boldsymbol{c}_i + \Delta \boldsymbol{c}_i \}_{i-1}^N.
\end{equation}
\end{small}

\smallskip

\noindent \textbf{Controllable 3D Editing.}
Next, we enable GDF to achieve controllable 3D editing to predict variable editing results.
Specifically, we apply the user-friendly slider as the output control interface for GDF.
In the implementation, our goal is to construct a learnable transformation $\mathcal{H}$ to transform the optimization trajectory from the original scene $\Theta$ to the target scene $\Theta_{edit}$ into a slider whose two endpoints correspond to a zero offset and a full offset~($\Theta_{edit} - \Theta$), respectively.
To achieve this goal, we extract the normalized control signal as input from the time step sequence of optimization.
For the intermediate result $\Theta_{edit}^t$, we utilize the ratio of $t$ to the total time steps $T$ as the control signal.
Considering that a larger total time step $T$ leads to dense and low-discriminative control signals, we discretize these signals into $k$ bins to reduce the learning difficulty of the transformation.
The inference process of GDF with control signals is shown in~\cref{eq:gdf3} and ~\cref{eq:gdf4}.
\begin{small}
\begin{equation}
\label{eq:gdf3}
(\Delta \boldsymbol{\mu}_i^t, \Delta \boldsymbol{\Sigma}_i^t, \Delta \sigma_i^t, \Delta \boldsymbol{c}_i^t) =
\begin{cases}
    {\mathcal{G}}(\boldsymbol{\mu}_i^0, \mathcal{H}(\frac{t}{T})), &\boldsymbol{\mathcal{M}}_i = 1, \\
    \overrightarrow{\mathbf{0}}, &\boldsymbol{\mathcal{M}}_i = 0,
\end{cases}
\end{equation}
\end{small}
\begin{small}
\begin{equation}
\label{eq:gdf4}
    \hat{\Theta}_{edit}^t = \{ \boldsymbol{\mu}_i^0 + \Delta \boldsymbol{\mu}_i^t, \boldsymbol{\Sigma}_i^0 + \Delta \boldsymbol{\Sigma}_i^t, \sigma_i^0 + \Delta \sigma_i^t, \boldsymbol{c}_i^0 + \Delta \boldsymbol{c}_i^t \}_{i-1}^N.
\end{equation}
\end{small}

\smallskip

\begin{figure*}[ht]
\hsize=\linewidth
\centering
\includegraphics[width=\linewidth]{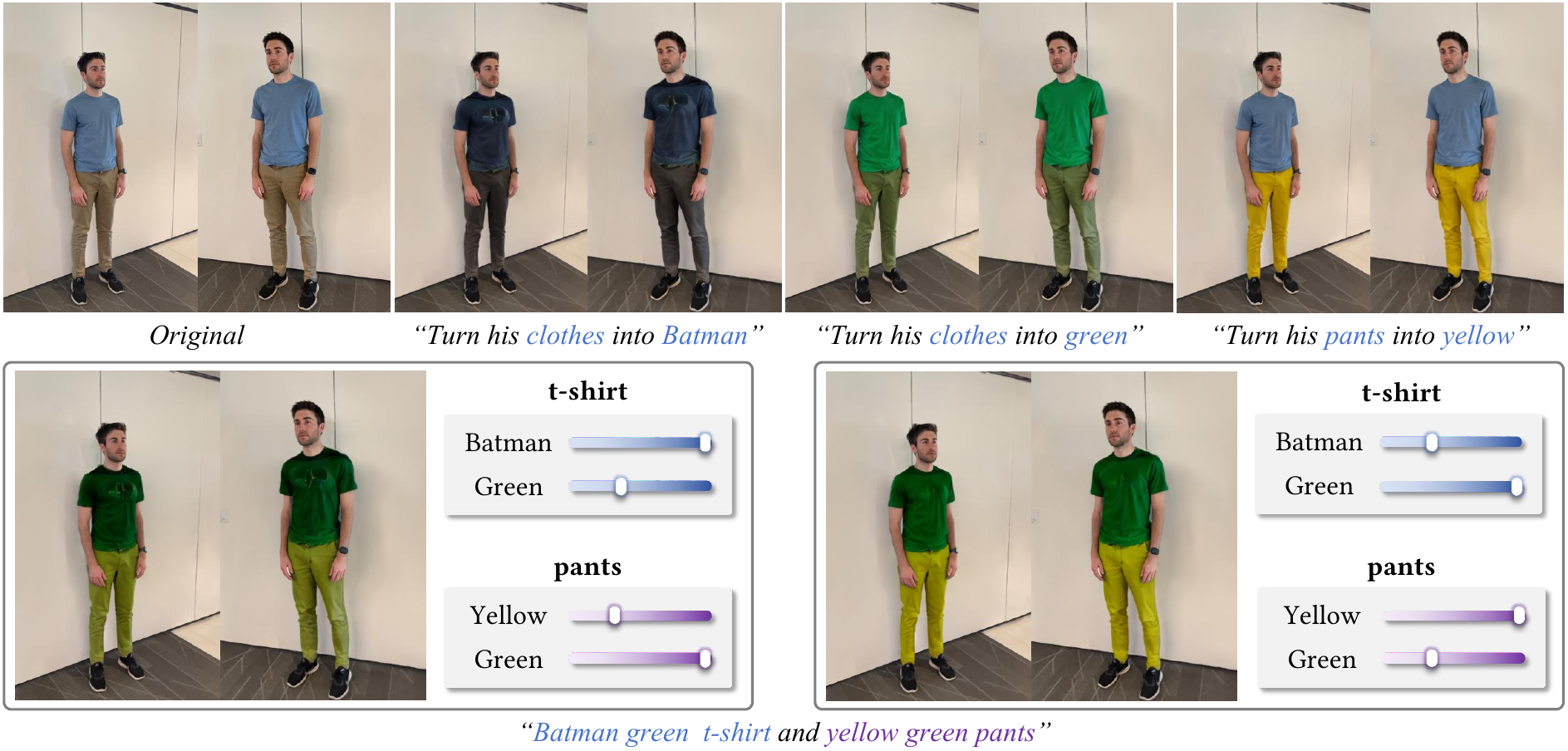}
\caption{
{Results of flexible fine-grained 3D manipulation.}
We model 3D editing processes as GDFs, and enable flexible fine-grained 3D manipulation by providing a specific manipulation region for each editing. 
}
\label{fig:manipulation}
\end{figure*}

\noindent \textbf{Trajectory Resampling.}
In addition, we propose the trajectory resampling strategy to prevent GDF from catastrophic forgetting during parallel training with 3D Gaussians due to sequential sampling of intermediate results as supervision.
Specifically, we sample the intermediate result uniformly according to time steps over the optimization trajectory and temporarily stores them in an online memory bank. 
Before each GDF iteration, we randomly sample an intermediate result from the memory bank for supervision.

\noindent \textbf{Loss Function.}
To optimize the GDF, we use the same editing loss as in~\cref{eq:loss_gdf_pred}. We sum the offsets predicted by GDF with the original 3D Gaussians to get the rendered images, and calculate the loss with the corresponding edited images.
\begin{equation}
\label{eq:loss_gdf_pred}
\begin{aligned}
    \mathcal{I}_{pred}^t = \mathcal{R}(\hat{\Theta}_{edit}^t, &v), \ 1 \leq t \leq T, \\
    \mathcal{L}_{pred} = \mathcal{L}_1(\mathcal{I}_{pred}^t, \mathcal{I}_r^t) &+ \mathcal{L}_{lpips}(\mathcal{I}_{pred}^t, \mathcal{I}_r^t),
\end{aligned}
\end{equation}
To improve the rendering quality, we also apply rendering quality regularization to GDF, \textit{cf.} \cref{eq:loss_gdf_render}, and the loss function of GDF is shown in~\cref{eq:loss_gdf}.
\begin{equation}
\label{eq:loss_gdf_render}
\mathcal{L}_{render} = -\nabla_{Laplacian}^2 \mathcal{R}(\hat{\Theta}_{edit}^t, v), \ 1 \leq t \leq T, 
\end{equation}
\begin{equation}
\label{eq:loss_gdf}
\mathcal{L}_{GDF} = \mathcal{L}_{pred} + \mathcal{L}_{render}.
\end{equation}

\section{Experiments}

\subsection{Main Results}
\noindent \textbf{Controllable 3D Editing.}
We first present the results of controllable 3D editing, \textit{cf.}~\cref{fig:controllable}.
Our method demonstrates superior controllable editing performance and provides variable results, which significantly facilitates the adjustment of 3D editing by providing users with a user-friendly slider.
For comparison, we present the results of Instruct-N2N \cite{haque2023instruct} (NeRF-based) and GaussianEditor \cite{chen2024gaussianeditor} (3DGS-based), \textit{cf.}~\cref{fig:comparison}.
For the same instructions, both provide only a single result in a single training session, neither allows for controllable editing.
In contrast, our method is capable of delivering variable and controllable results in a single training session.
Moreover, the editing results of our method are also superior to existing methods.

\noindent \textbf{Flexible Fine-grained 3D Manipulation.}
Next, we present the editing results of flexible fine-grained 3D manipulation, \textit{cf.}~\cref{fig:manipulation}.
We model different 3D editing processes as corresponding GDFs, and implement fine-grained 3D manipulation by providing a specific editing region for each GDF.
The editing result of ``Batman green t-shirt and yellow green pants'' is combined by three results in the first row, where the results with overlapping editing regions are able to control the main theme using the slider provided by controllable editing.
All implementation details of ProGDF and more results on two applications are shown in the Appendix to demonstrate the extensive applicability of our method.

\noindent \textbf{Quantitative Results.}
We perform a quantitative comparison on user study and CLIP directional similarity~($S_{dir}$) following~\cite{chen2024gaussianeditor}~(see the Appendix for the user study evaluation criteria). Due to the unique advantages of ProGDF, we randomly adjust its slider for three inference runs and take the best results to compare with Instruct-N2N and GaussianEditor. Our method achieves the superior performance for both user study and CLIP $S_{dir}$, \textit{cf.}~\cref{tab:quantitative}.

\begin{table}[t]
\centering
\renewcommand{\arraystretch}{1.15}
\setlength{\tabcolsep}{1.6mm}
\footnotesize
\begin{tabular*}{\linewidth}{l | ccc}
\noalign{\hrule height 1.2pt}
Metric & Instruct-N2N & GaussianEditor & ProGDF (Ours) \\
\hline
User study $ \uparrow $ & 2.84 $\pm$ 0.20  & 3.32 $\pm$ 0.40 & \textbf{4.06 $\pm$ 0.22} \\
CLIP $S_{dir}$ $ \uparrow $ & 0.1600 & 0.2071 & \textbf{0.2180} \\
\noalign{\hrule height 1.2pt}
\end{tabular*}
\caption{Quantitative results. $S_{dir}$ is directional similarity.}
\vspace{-2mm}
\label{tab:quantitative}
\end{table}

\subsection{Ablations}

\begin{figure}[ht]
\hsize=\linewidth
\centering
\includegraphics[width=\linewidth]{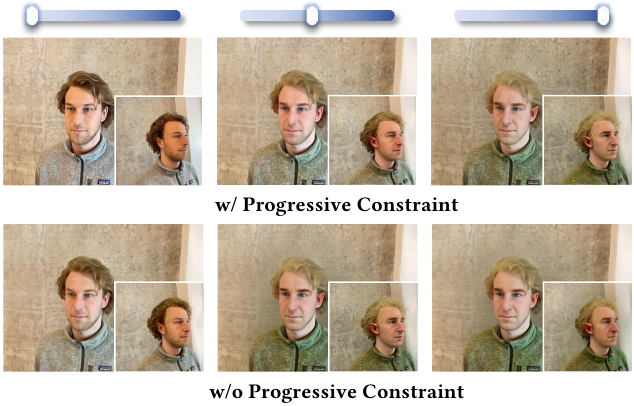}
\caption{
{Ablation on progressive constraint.} Removing this constraint leads to a lack of distinctiveness in the results.
}
\label{fig:ablation_pc}
\end{figure}
\noindent \textbf{Progressive Constraint.}
To verify the role of the progressive constraint, we remove it to train the GDF and then inference with different control signals to generate different editing results for comparison, as shown in~\cref{fig:ablation_pc}.
Due to the removal of the progressive constraint, the 3D Gaussians converge to the target result directly under the guidance of $\mathcal{L}_{edit}$ and $\mathcal{L}_{render}$, lacking various intermediate results.
As a result, the GDF generates similar results in terms of texture detail and color saturation for different control signals: the last two results for ``w/o progressive constraints'' are very similar, while the results for ``w/ progressive constraints'' exhibit significant distinctions.

\begin{figure}[ht]
\hsize=\linewidth
\centering
\includegraphics[width=\linewidth]{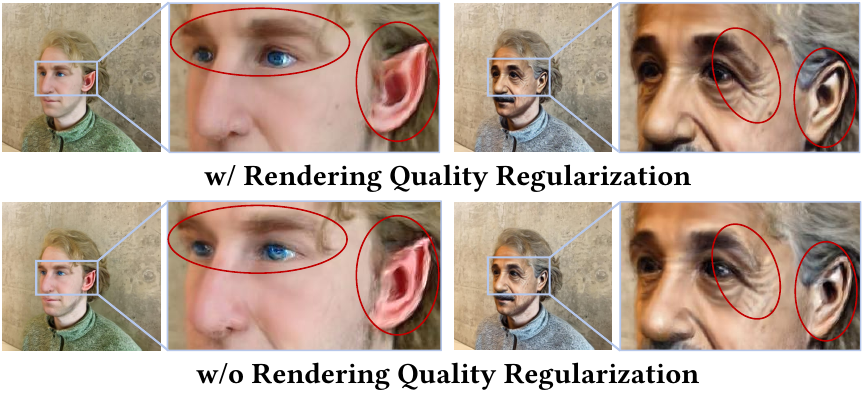}
\caption{
{Ablation on rendering quality regularization.} Removing this regularization leads to low-quality rendering results~(marked by red circles).
}
\label{fig:ablation_rq}
\end{figure}
\noindent \textbf{Rendering Quality Regularization.}
To verify the role of the rendering quality regularization, we remove it from the loss functions of PGS and GDF to generate editing results for comparison.
Removing the rendering quality regularization causes texture blurring in the rendering results, producing lower quality edits, such as the ears, eyes and wrinkles marked with red circles in~\cref{fig:ablation_rq}.
\section{Related Work}

\subsection{2D Editing}
Recently, many diffusion-based image editing methods \cite{meng2021sdedit,hertz2022prompt,kawar2023imagic,cao2023masactrl,brooks2023instructpix2pix,ruiz2023dreambooth} have been proposed inspired by the impressive results of large-scale text-to-image diffusion models~\cite{rombach2022high,saharia2022photorealistic}.
Although some text-to-image models~\cite{ramesh2022hierarchical} possess their inherent image editing capabilities, it is non-trivial to achieve the desired editing due to their inability to overcome the intrinsic randomness of the generated content.
To address this issue, Prompt-to-Prompt~\cite{hertz2022prompt} establishes a relationship between the image spatial layout and each word in the prompt to enable flexible editing of the generated image.
To facilitate the editing of realistic images, SDEdit~\cite{meng2021sdedit} achieves the balance between realism and faithfulness by iteratively denoising the noised images through a stochastic differential equation.
Inspired by research on teaching large language models to better follow human instructions for linguistic tasks~\cite{mishra2021cross,ouyang2022training,wei2021finetuned}, instruction-following image editing methods have been proposed~\cite{brooks2023instructpix2pix}.
The user simply provides an input image and a brief editing instruction for the desired editing, which is more expressive, precise and intuitive.
To achieve more scalable conditional control in text-to-image, ControlNet~\cite{zhang2023adding} and T2I-Adapter~\cite{mou2024t2i} introduce additional trainable modules for guided image generation, which has gained much attention from the community.

\subsection{3D Editing}
\noindent \textbf{NeRF-based.} The complexity of the scene data encoding process makes Neural Radiance Fields~(NeRF)~\cite{mildenhall2021nerf} challenging to edit.
EditNeRF~\cite{liu2021editing} represents a pioneering contribution to the field, as it enables the editing of the shape and color of neural fields by conditioning them on latent codes.
Some works~\cite{wang2022clip,gao2023textdeformer} employ the CLIP~\cite{radford2021learning} model to facilitate editing through the utilization of text prompts or reference images.
In addition to color editing, some works~\cite{yang2022neumesh,xu2022deforming,yuan2022nerf} transform implicit neural fields into explicit meshes and establish correspondences to perform controlled shape deformation.
Recently, some works~\cite{haque2023instruct,zhuang2023dreameditor,liu2024genn2n} employ 2D image editing models for NeRF editing and achieve impressive results.
However, these NeRF-based editing methods are limited by the complexity of implicit neural fields and bottlenecks in training and rendering.

\smallskip

\noindent \textbf{3DGS-based.} The inherently explicit nature of 3DGS makes it easy to manipulate specific regions, providing a distinct advantage in editing tasks.
GaussianEditor~\cite{chen2024gaussianeditor} employs the capacity of 2D editing models to optimize Gaussians within specific regions of the scene for text-driven editing.
GSEdit~\cite{palandra2024gsedit} incorporates post-processing for mesh extraction and texture refinement, following a similar editing procedure.
\cite{wang2024gaussianeditor} provides a more automated 3D editing pipeline in combination with LLM.
In addition, StyleGaussian~\cite{liu2024stylegaussian} enables efficient 3D style transfer with 3DGS.
However, existing 3D editing methods all adopt the paradigm of in-the-loop training, which ignore the controllability and flexibility of the results, resulting in users spending tens of minutes to optimize a large number of 3D Gaussians for each adjustment, at high computational cost and time overhead.
Our method is an out-of-loop training approach that requires only a single training session to provide users with controllable editing capability and variable editing results through a user-friendly interface in real-time.
\section{Limitations}
Although our ProGDF elegantly achieves controllable and flexible 3D editing, there are a few limitations that need to be addressed. These include the need for exploring alternative interactive types for 3D editing~(\textit{e.g.}, drag, click, or scribble, \textit{etc.}) beyond the basic slider we currently provide. Additionally, we have yet to explore decoupled control of texture and color for 3D Gaussians to further improve the controllability and flexibility of 3D editing. These limitations will be the focus of our future research.

\section{Conclusion}
In this work, we introduce the concept of process-oriented modeling for 3D editing and propose the Progressive Gaussian Differential Field (ProGDF), which is the first out-of-loop training approach for 3D editing.
Our approach requires only a single training session to provide users with controllable editing capability and variable editing results through a user-friendly interface in real-time.
We present two novel applications, namely controllable 3D editing and flexible fine-grained 3D manipulation, to showcase the effectiveness and versatility of our method. 
Through our research, we aim to provide valuable insights into the practice of 3D editing.

\bibliography{aaai25}
\clearpage
\renewcommand{\thefootnote}{\fnsymbol{footnote}}

\renewcommand{\thetable}{\Alph{table}}
\renewcommand{\theequation}{\Alph{equation}}
\renewcommand{\thefigure}{\Alph{figure}}

\setcounter{table}{0}
\setcounter{section}{0}
\setcounter{figure}{0}
\setcounter{equation}{0}

\maketitleappendix

\begin{figure*}[!h]
\hsize=\linewidth
\centering
\includegraphics[width=\linewidth]{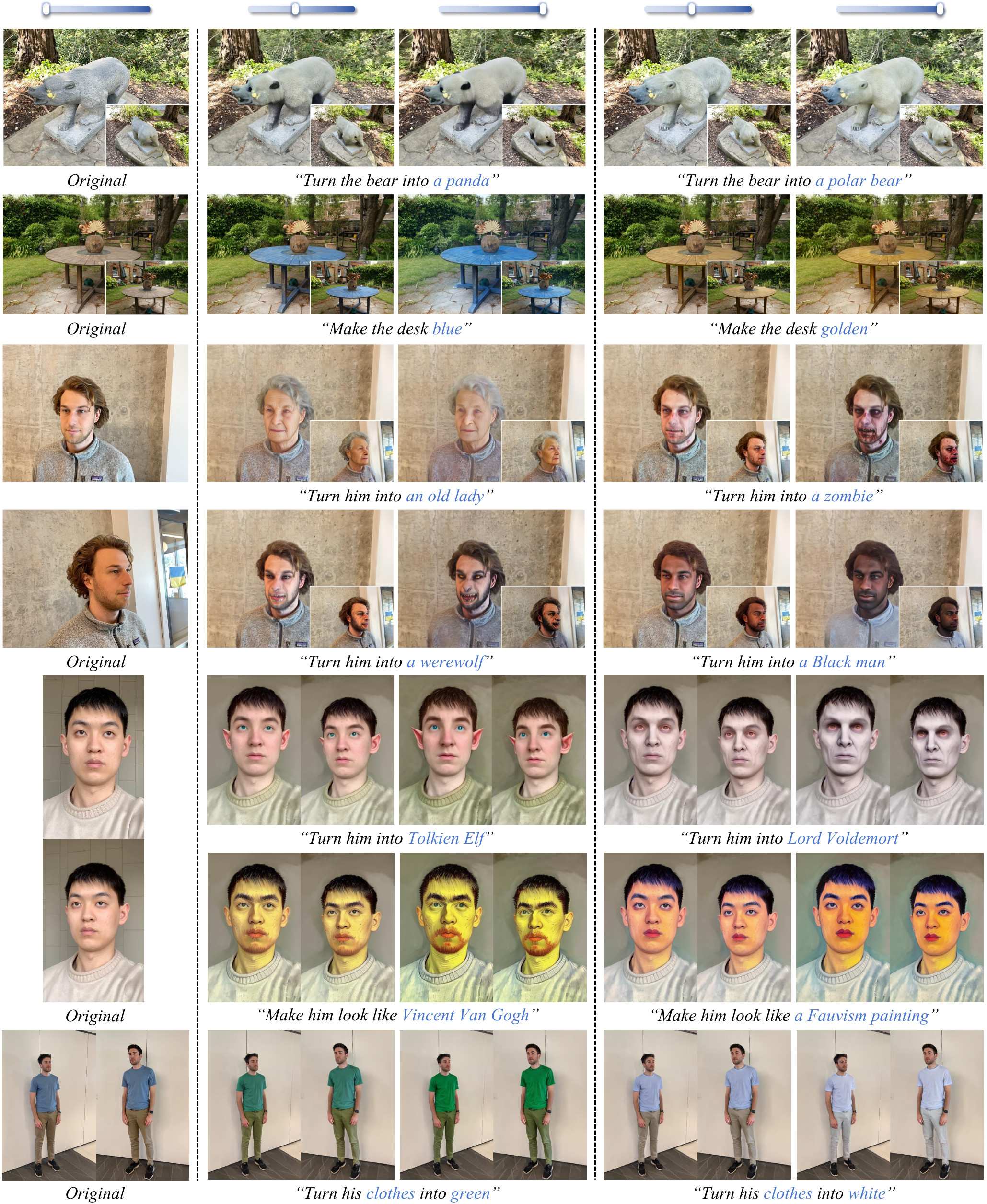}
\caption{
Extensive results of controllable 3D editing.
}
\label{fig:supp_result1}
\end{figure*}

\begin{figure*}[!h]
\hsize=\linewidth
\centering
\includegraphics[width=\linewidth]{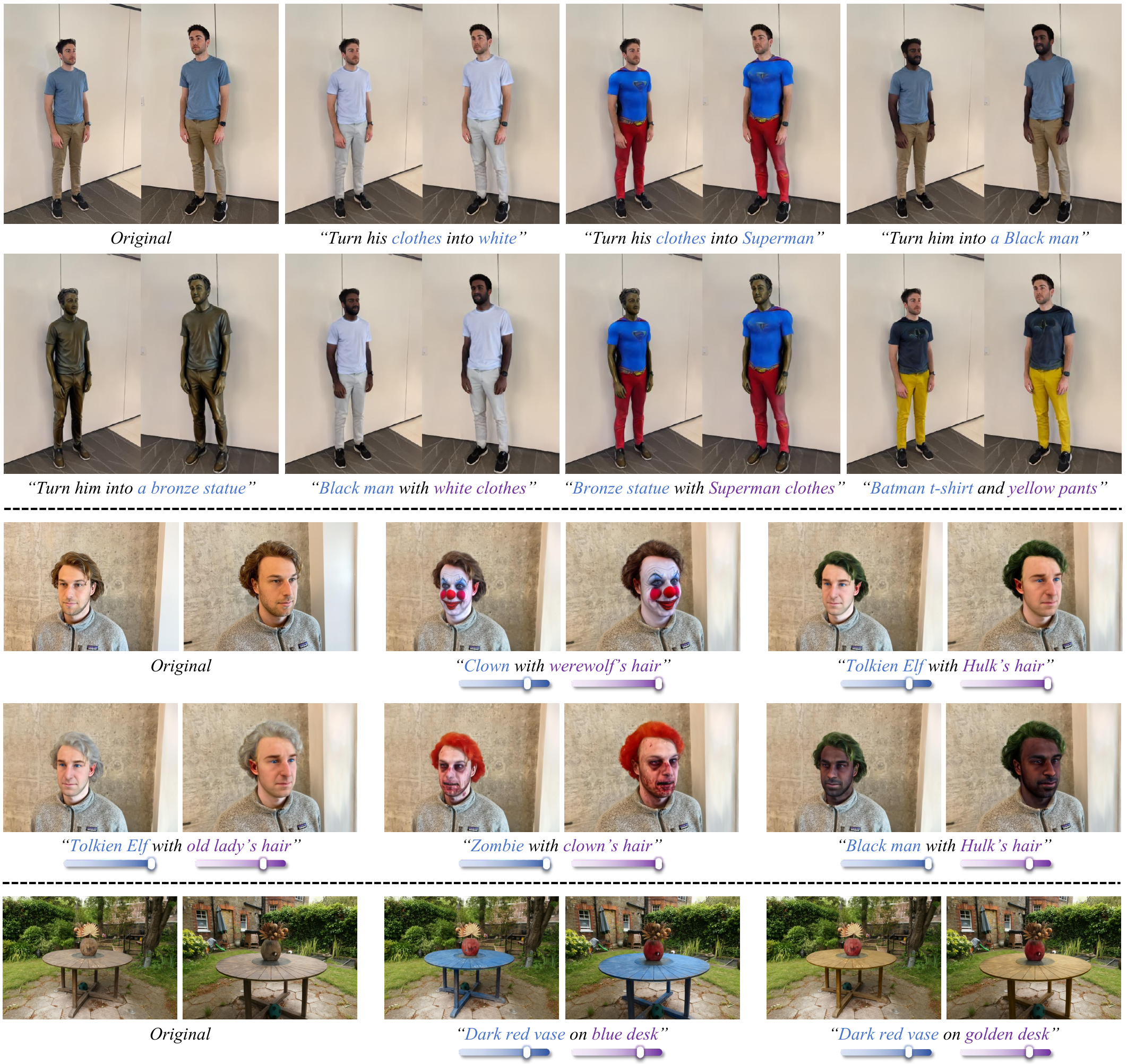}
\caption{
Extensive results of flexible fine-grained 3D manipulation.
}
\label{fig:supp_result2}
\end{figure*}

\section{Preliminary: 3D Gaussian Splatting}
Gaussian Splatting~\cite{kerbl20233d} provides a real-time differentiable renderer and explicitly represents a 3D scene as a set of 3D Gaussian primitives $ \Theta = \{ (\boldsymbol{\mu}_i, \boldsymbol{\Sigma}_i, \sigma_i, \boldsymbol{c}_i) \}_{i=1}^N $.
In this representation, each 3D Gaussian $\Theta_i$ is parameterized by a center point $\boldsymbol{\mu}_i$ and a covariance matrix $\boldsymbol{\Sigma}_i$, representing the distribution as:
\begin{equation}
\label{equ:gs}
    \Theta_i (\boldsymbol{x}) = e^{-\frac{1}{2} \textit{$\boldsymbol{x}$}^{T} \boldsymbol{\Sigma}_i^{-1} \textit{$\boldsymbol{x}$} }.
\end{equation}
To introduce spatial variations in blending across the scene, an opacity $\sigma_i$  is employed to control the influence of each Gaussian.
To obtain a physically meaningful covariance matrix, which must be positive semi-definite, the following equivalent representation is adopted:
\begin{equation}
    \boldsymbol{\Sigma}_i = \boldsymbol{R}_i\boldsymbol{S}_i\boldsymbol{S}_i^{T}\boldsymbol{R}_i^{T},
\end{equation}
where the scaling and rotation matrices are represented as a scaling factor $\boldsymbol{S}_i$ and a rotation quaternion $\boldsymbol{R}_i$.
Each Gaussian represents its color with spherical harmonics coefficients, and we omit view-dependency for simplicity and utilize $\boldsymbol{c}_i$ to encode appearance.

In summary, each 3D Gaussian is parameterized by a set of attributes: position $\boldsymbol{\mu}_i \in \mathbb{R}^3$, scaling factor $\boldsymbol{S}_i \in \mathbb{R}^3$, rotation quaternion $\boldsymbol{R}_i \in \mathbb{R}^4$, opacity $\sigma_i \in \mathbb{R}$, and color $\boldsymbol{c}_i \in \mathbb{R}^3$.
Practically, 3D Gaussians can be effectively rendered to compute the color $\boldsymbol{C}$ by blending $N$ ordered Gaussians overlapping the pixel:
\begin{equation}
\label{equ:volume_render}
    \boldsymbol{C} = \sum_{i \in N} \boldsymbol{c}_i \alpha_i \prod_{j=1}^{i-1} (1-\alpha_j),
\end{equation}
where $\alpha_i$ is calculated by evaluating $\Theta_i$ with~\cref{equ:gs} multiplied by its opacity $\sigma_i$. 

\section{Implementation Details}
All of the original 3D Gaussians we used are trained using the method described in~\cite{kerbl20233d}, with raw data from the MIP-NeRF and Instruct-N2N datasets, and rendering during training is performed using the highly optimized renderer proposed in~\cite{kerbl20233d}.
For region-specific editing, we utilize SAM~\cite{kirillov2023segment} to segment multiple views and calculate the 3D mask follow~\cite{chen2024gaussianeditor}.
The experiment uses a maximum of $96$ viewpoints.
We employ InstructPix2Pix~\cite{brooks2023instructpix2pix} as the image editing model in our framework.
For progressive constraint, we set the initial gradient weight $\alpha$ to $0.05$, the decay coefficient $\beta$ to $1.1$, and $s$ to $50$.
We discretize the control signal of the GDF into $k = 10$ bins~\cite{zhao2024graco}.
For trajectory resampling, the sampling interval is set to $100$ by default.
We use PyTorch for implementation and a single 32GB NVIDIA V100 GPU for training and inference.
Following the previous works~\cite{chen2024gaussianeditor,palandra2024gsedit}, we set the total time steps $T$ to 1500-2000, which varies depending on the editing instruction and the complexity of the scene, taking about tens of minutes.
Our GDF is lightweight~($\sim$6M) and can be trained in parallel with 3D Gaussians, so the additional computing and storage costs are acceptable.
After training, the inference time of the GDF is only $0.02$ seconds.

\section{Manipulation Region Extraction}
To achieve fine-grained 3D manipulation, we extract different manipulation regions from the 3D Gaussians and assign them to different GDFs.
Specifically, we assign a binarized label to each 3D Gaussian, where $1$ indicates that it is inside the selected region and $0$ indicates that it is outside.
We utilize $\boldsymbol{\mathcal{M}}$ to represent a specific manipulation region of a 3D scene consisting of the set of 3D Gaussians $\{ \Theta_i \ | \ i \in N, \boldsymbol{\mathcal{M}}_i=1 \}$.

To obtain the target manipulation region $\boldsymbol{\mathcal{M}}$, we perform 2D segmentation for multi-view images using LangSAM, a variant of SAM~\cite{kirillov2023segment} that supports text-prompted segmentation, which is capable of generating 2D masks of all viewpoints based on user-provided text prompt.
Then, we unproject these 2D masks containing the target object into the 3D space to obtain the label for each Gaussian.
Formally, for a set of 3D Gaussians $\Theta$, we first employ the differentiable renderer~\cite{kerbl20233d} to obtain a collection of multi-view images $\boldsymbol{I}_r$ and then employ LangSAM to perform 2D segmentation to obtain a series of 2D masks $\boldsymbol{m}$.
We unproject the 2D masks $\boldsymbol{m}$ onto the 3D Gaussians $\Theta$ by inverse rendering, which is calculated as~\cref{equ:inverse}.
\begin{equation}
\label{equ:inverse}
w_i = \sum_{p} \sigma_i \cdot \alpha_i \prod_{j=1}^{i-1} (1-\alpha_j) \cdot \boldsymbol{m}(p),
\end{equation}
where $w_i$ represents the weight of the $i$-th Gaussian $\Theta_i$, $\sigma_i$ represents the opacity of $\Theta_i$, $\alpha_i$ is the same as~\cref{equ:volume_render}, and $\boldsymbol{m}(p)$ denotes the label of pixel $p$.
Meanwhile, we count the number of times the weights are accumulated for each Gaussian $n_i$, which indicates how many pixels corresponding to rays pass through the Gaussian $\Theta_i$. Finally, we average the weights of each Gaussian with $\frac{w_i}{n_i}$.
We determine whether a Gaussian $\Theta_i$ belongs to the target manipulation region by whether its average weight exceeds a predefined threshold $\epsilon$~(set to $0.8$ in our experiments), \textit{cf.}~\cref{eq:mask}.
\begin{equation}
\label{eq:mask}
\boldsymbol{\mathcal{M}}_i = 
\begin{cases} 
1, &  \frac{w_i}{n_i} \geq \epsilon, \\
0, &  \frac{w_i}{n_i} < \epsilon.
\end{cases}
\end{equation}

\section{User Study}
We conduct user study and report results in the main paper, see~\cref{tab:study} for detailed evaluation criteria.
Specifically, we ask the participants to rate along three dimensions: \textit{accuracy} of understanding the instructions, \textit{rationality} of the editing results, and \textit{quality} of the editing results.
All dimensions are scored on a $5$-point level, and we ultimately take the average of the scores across these dimensions as the user study score and provide the 95\% confidence interval.
The user study results are from a total of $30$ participants.

\begin{table*}[t]
\renewcommand{\arraystretch}{1.15}
\begin{tabular*}{\linewidth}{l | c | p{13.94cm}}
\noalign{\hrule height 1.2pt}
\textbf{Dimension} & \textbf{\#Point} & \textbf{Description} \\
\hline
\multirow{5}{*}[-4.7ex]{\textbf{Accuracy}} & 1 & Very poor, the system barely understands the instructions and does not match the user's intent at all. \\
\cline{2-3}
& 2 & Rather poor, the understanding of the instructions is highly inaccurate and does not meet the user's intent. \\
\cline{2-3}
& 3 & Acceptable, the understanding of the instructions is basically correct and generally meets the user's intent. \\
\cline{2-3}
& 4 & Fairly good, the understanding of the instructions is relatively accurate and meets the user's intent, but there are still shortcomings. \\
\cline{2-3}
& 5 & Very good, the system understands the instructions very accurately and there are no obvious shortcomings. \\
\noalign{\hrule height 0.9pt}
\multirow{5}{*}[-4.7ex]{\textbf{Rationality}} & 1 & Very poor, the result is not reasonable at all, there is severe distortion or the original features are completely lost. \\
\cline{2-3}
& 2 & Rather poor, the result is rather unreasonable, there is significant distortion or very little of the original features are retained. \\
\cline{2-3}
& 3 & Acceptable, the result is basically reasonable, there is some distortion, and the original features are generally recognizable. \\
\cline{2-3}
& 4 & Fairly good, the result is reasonable, there is some distortion, the original features are recognizable. \\
\cline{2-3}
& 5 & Very good, the result is clearly reasonable, there is no obvious distortion and the original features are fully recognizable. \\
\noalign{\hrule height 0.9pt}
\multirow{5}{*}{\textbf{Quality}} & 1 & Very poor, texture detail is very blurred, color distribution anomalous. \\
\cline{2-3}
& 2 & Rather poor, texture detail is blurred, color distribution is sometimes anomalous. \\
\cline{2-3}
& 3 & Acceptable, texture detail is slightly blurred, color distribution is basically normal. \\
\cline{2-3}
& 4 & Fairly good, texture detail is relatively clear, color distribution is normal. \\
\cline{2-3}
& 5 & Very good, texture detail is very clear, color distribution is very reasonable. \\
\noalign{\hrule height 1.2pt}
\end{tabular*}
\caption{{The detailed evaluation criteria of the user study.}}
\label{tab:study}
\end{table*}

\section{Additional Results}
We present more extensive results of controllable 3D editing, \textit{cf.}~\cref{fig:supp_result1}.
We provide more editing instructions on different 3D scenes to illustrate the effectiveness and controllability of the proposed ProGDF.
We also present extensive results of flexible fine-grained 3D manipulation, \textit{cf.}~\cref{fig:supp_result2}.
These results highlight the advantages of our proposed process-oriented modelling for 3D editing, greatly improving the user experience of the 3D editing tools.

\end{document}